# Disorder driven quantum critical behavior in CuGeO$_3$ doped with magnetic impurity

S.V. Demishev [a,*], A.V. Semeno [a], N.E. Sluchanko [a], N.A. Samarin [a], A.A. Pronin [a], V.V. Glushkov [a], H. Ohta [b], S. Okubo [b], M. Kimata [b], K. Koyama [c], M. Motokawa [c], A.V. Kuznetsov [d]

[a] *General Physics Institute of RAS, Vavilov st.,38, 119991 Moscow, Russia*

[b] *Molecular Photoscience Research Center, Kobe University, 1-1 Rokkodai, Nada, Kobe, 657-8501, Japan*

[c] *Institute for Materials Research, Tohoku University, Sendai 980-8577, Japan*

[d] *Moscow Engineering Physics Institute, Kashirskoe Shosse, 31, 115409 Moscow, Russia*

**Abstract**

For the CuGeO$_3$ doped with 1% of Fe the quantum critical behavior in a wide temperature range 1-40 K is reported. The critical exponents for susceptibility along different crystallographic axes are determined: $\alpha$=0.34 (**B**||**a** and **B**||**c**) and $\alpha$=0.31 (**B**||**b**). New effect of the frequency dependence of the critical exponent is discussed.

Recently we have shown that doping with 1% of Fe impurity (S=2) of CuGeO$_3$ induces in high quality single crystals a complete damping of both spin-Peierls and Neel transitions and leads to a quantum critical (QC) behavior [1-4]. This result has been experimentally confirmed by magnetic susceptibility and specific heat data as well as by studying of the magnetic phase diagram [1-4]. Details about samples preparation and characterisation can be found elsewhere [1]. In the QC regime for $T<T_G$ the ground state is represented by Griffiths phase (GP) [1-5]. The fingerprint of the GP is a divergent magnetic susceptibility, which acquires essentially non Curie-Weiss form,

$$\chi(T) \sim 1/T^{\alpha}, \qquad (1)$$

where $\alpha<1$ [1-5]. The power law given by equation (1) have been reported for the limited temperature range 1.8-30 K with the critical index $\alpha$=0.36 and [1-4]. Therefore the question about temperature asymptotic of magnetic susceptibility at very low temperatures, which is essential for the quantum critical problem, remains open. The second unconsidered up to now predicament is that the critical behaviour given by equation (1) have been checked only for the case when external magnetic field **B** was parallel to **a** crystallographic axis [1-4] and the information for **B**||**b** and **B**||**c** is missing. In the present work we are aimed on getting experimental answers on the two aforementioned problems.

Spin part of the magnetic susceptibility (1) have been probed by 60-100 GHz cavity EPR spectrometers and for 100-315 GHz by quasi-optical technique. For the 60 GHz cavity we were able to extend EPR measurements down to 0.5 K using He$^3$ cryostat. For the absolute calibration of $\chi(T)$ the independent measurements of the same crystal have been performed in SQUID magnetometer. The studied single crystals of CuGeO$_3$ contained 1% of Fe impurity.

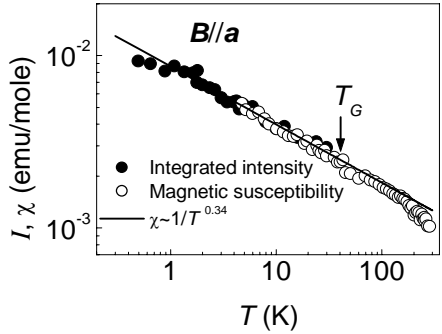

Fig. 1. Temperature dependence of the integrated intensity at 60 GHz and magnetic susceptibility.

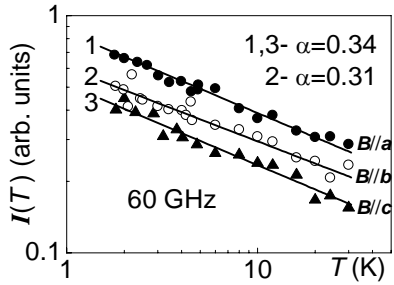

Fig. 2. The $I(T)$ data for $B\|a$, $B\|b$ and $B\|c$.

The experimental EPR spectra for $CuGeO_3$:Fe consist of a single line corresponding to resonance on S=1/2 chains of $Cu^{2+}$ ions in agreement with previous findings [1-4]. A full set of spectroscopic data (line widths, $g$-factors, integrated intensities) for the whole temperature-frequency domain studied was obtained. Here we will restrict ourselves with the integrated intensity $I(T)$; the rest of the data will be published elsewhere.

It is found that power law (1) for the integrated intensity $I(T) \sim \chi(T)$ is fulfilled in the temperature range $1 < T < 40$ K (fig. 1). The observed critical exponent is 0.34±0.02 and practically coincide with the published results [1-4]. Below 1 K the $I(T)$ curve tends to saturate, but no sign of antiferromagnetic transition was found down to the lowest temperature studied. The beginning of the power law provides an estimate of the temperature for the transition into Griffiths phase $T_G \sim 40$ K

Considering anisotropy of the $I(T)$ we found that for $B\|a$ and $B\|c$ the data correspond to α=0.34±0.02 and for $B\|b$ the index is smaller: α=0.31±0.02 (fig. 2). Nevertheless all observed values are very close and thus strongly supports idea

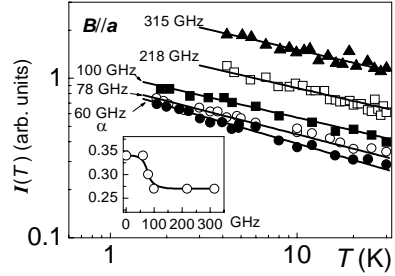

Fig. 3. Frequency dependent QC behavior.

about universal behaviour of susceptibility in the quantum critical regime [1-5].

Interesting that in the low frequency range α=0.34 whereas for high frequencies the index is about α=0.27. The data in fig. 3 suggest that transition between low and high frequency asymptotics occur in the vicinity of 78-100 GHz.

In conclusion, we have shown that critical behaviour described by equation (1) can be observed in $CuGeO_3$:Fe in a wide range where temperature vary 40 times. The critical exponents along three crystallographic axes are very close that is in agreement with the QC model. The reasons for the frequency dependence of the critical index are not clear and may probably reflect the dynamical properties of the spin clusters in the Griffiths phase.

Authors acknowledge support from Russian Science Support Foundation, programmes "Physics of Nanostructures" and "Strongly Correlated Electrons" of RAS and grants RFBR 04-02-16574, INTAS 03-51-3036 and Grant-in-Aid No. 13130204.